# Quenched lattice fluctuations in optically driven SrTiO$_3$

M. Fechner[1], M. Först[1], G. Orenstein[2], V. Krapivin[2], A.S. Disa[1], M. Buzzi[1], A. von Hoegen[1], G. de la Pena[2], Q. L Nguyen[2,3], R. Mankowsky[4], M. Sander[4], H. Lemke[4], Y. Deng[4], M. Trigo[2] and A. Cavalleri[1,5]

[1]Max Planck Institute for the Structure and Dynamics of Matter, Hamburg, Germany
[2]Stanford Pulse Institute, SLAC National Accelerator Laboratory, Menlo Park, CA, USA
[3]Linac Coherent Light Source, SLAC National Accelerator Laboratory, Menlo Park, CA, USA
[4]Swiss Light Source, Paul Scherrer Institut, Villigen, Switzerland
[5]Department of Physics, Clarendon Laboratory, University of Oxford, Oxford, United Kingdom

**Many functionally relevant ferroic phenomena in quantum materials can be manipulated by driving the lattice coherently with optical and terahertz pulses[1–13]. New physical phenomena and non-equilibrium phases that have no equilibrium counterpart have been discovered following these protocols. The underlying structural dynamics has been mostly studied by recording the average atomic position along dynamical structural coordinates with elastic scattering methods[14–21]. However, crystal lattice fluctuations, which are known to influence phase transitions in equilibrium, are also expected to determine these dynamics but have rarely been explored[22]. Here, we study the driven dynamics of the quantum paraelectric SrTiO$_3$, in which mid-infrared drives have been shown to induce a metastable ferroelectric state[9]. Crucial in these physics is the competition between the polar instability and antiferrodistortive rotations, which in equilibrium frustrate the formation of long-range ferroelectricity. We make use of high intensity mid-infrared optical pulses to resonantly drive a Ti-O stretching mode at 17 THz, and we measure the resulting change in lattice fluctuations using time-resolved x-ray diffuse scattering at a free electron laser. After a prompt increase, we observe a long-lived quench in *R-point* antiferrodistortive lattice fluctuations. The enhancement and reduction in lattice fluctuations are explained theoretically by considering fourth-order nonlinear phononic interactions and third-order coupling to the driven optical phonon and to lattice strain, respectively. These observations provide a number of new and testable hypotheses for the physics of light-induced ferroelectricity.**

The cubic perovskite structure is inherently unstable, frequently undergoing different types of symmetry-breaking structural distortions. SrTiO$_3$ (STO) is an especially interesting case, in which both ferroelectric (FE) and anti-ferrodistortive (AFD) instabilities (Fig. 1a,b) compete to determine the ground state. Above 110 K, the cubic structure appears to be primed for a ferroelectric transition, which one would expect upon cooling as a polar distortion is the dominant symmetry breaking for this lattice structure. However, ferroelectricity fails to materialize, because below 110 K an anti-ferrodistortive instability induces a low-symmetry tetragonal phase. Although large polar fluctuations set in below 30 K, resembling a precursor for the ferroelectric phase transition, long range polar order is stifled by large zero-point fluctuations. This phenomenon is sometimes referred to as quantum paraelectricity[23,24]. This frustrated state of matter can be lifted by either isotope-replacement of oxygen[25], partial cation doping [26], or epitaxial strain[27], which trigger the formation of a macroscopic ferroelectric polarization.

These physics are discussed in quantitative detail[28] in Figure 1c,d, where we report the results of *ab-initio* calculations. In the cubic structure (rotation angle $\phi_{AFD} = 0$), the potential energy along the ferroelectric coordinate $P$ exhibits a distinct minimum at a finite polarization $P_0$, indicating the existence of a ferroelectric instability. When the tetragonal distortion and the antiferrodistortive rotations set in, with a calculated equilibrium value of $\phi_{AFD} = 6°$, the potential minimum along the $P$ becomes significantly shallower. Here, the condensation into the ferroelectric phase becomes suppressed, as the depth of potential energy at $P_0$ is comparable to the amplitude of zero-point fluctuations of the ferroelectric soft mode. A further increase in AFD rotation to $\phi = 12°$ would revive the deeper ferroelectric instability, with a potential minimum similar in size to that of the cubic state at $\phi = 0°$. Hence, the quantum paraelectric state of STO with suppressed ferroelectricity exists only in the low-temperature tetragonal structure for a narrow range of AFD rotation angles near $\phi = 6°$. Note that experimentally, the absolute rotation angle may be slightly different from $\phi = 6°$, although the physical picture remains the same.

Intense terahertz (THz) and mid-infrared light pulses were shown to remove this frustration and to induce ferroelectricity [9,10], either transiently when coupling to the soft mode directly (Ref. 10) or permanently when a higher-frequency auxiliary mode was driven (Ref. 9). In the latter case, it was argued that the coupling between the resonantly driven Ti-O stretch phonon to acoustic modes may produce a strain field that stabilizes a long-range polar phase. Although the dynamics of the antiferrodistortive mode are also expected to contribute, their role in these physics has not been established.

Here, we used time-resolved x-ray diffuse scattering[29] to experimentally map the lattice fluctuations in the cubic high-temperature phase of STO, measured at the Brillouin zone boundary (½ ½ ½) *R-point*. These experiments were performed on a (100)-oriented STO substrate at the Bernina end station of the SwissFEL[30] free electron laser. The sample was cooled by a cryogenic nitrogen gas jet to approximately 135 K, i.e. above the AFD structural transition, and was excited by mid-infrared (mid-IR) pulses of ~150 fs duration with central frequency and bandwidth of 18 THz and 5 THz full width at half maximum, respectively, and fluences up to 60 mJ/cm$^2$. This pump is resonant with the highest-frequency infrared-active STO phonon at 17 THz and is polarized nearly parallel to the [001] crystallographic direction. The x-ray probe pulses (time duration ~ 50 fs) were tuned to 9.0 keV photon energy and spectrally filtered to ~1 eV by a Si (111) monochromator before being focused on the sample. The probe beam was sent through a hole bored into an off-axis parabola used to focus the mid-IR beam, enabling collinear excitation and x-ray probing (see Fig. 2a). The temporal jitter between pump and probe pulses was monitored using a spectral encoding technique on a shot-by-shot basis and corrected in the post-processing.

The diffracted x-rays were detected by a Jungfrau pixel array detector, positioned 100 mm from the sample, and normalized to the incident x-ray intensity. Figure 2b shows a typical detector image integrated over 1,000 x-ray pulses, which shows localized diffuse scattering at the *R-point* (1.5 2.5 3.5) and a broader feature around the *M-point* (2.0 2.5 3.5). We chose a grazing exit geometry with an exit angle of about 1.5° for the *R-point* scattering to reduce the escape depth of the diffracted x-ray probe to nearly match the penetration depth of the mid-IR pump while simultaneously allowing for high excitation fluence at near-normal incidence. The left part of the image, separated by the relatively sharp sample horizon, results from air scattering.

Away from Bragg reflections, as is the case here, the diffuse scattering intensity $I(q)$ at a reduced wavevector $q = K - G$ is proportional to the variance of the atomic displacements $\langle u_q u_{-q} \rangle$, connected to this wavevector[31]. Here, $K$ is the total momentum transfer of the scattering process and $G$ the nearest reciprocal lattice vector. At a structural phase transition, the divergence of lattice fluctuations generally results in a critical increase of the diffuse scattering intensity[32]. In the particular case of STO, the cubic-to-tetragonal phase transition at 110 K is driven by a softening and condensation of the AFD phonon mode at the *R-point* in reciprocal space[33]. The increase of lattice fluctuations associated with the onset of this structural transition generates the enhanced scattering intensity at the well-defined detector region corresponding to the *R-point*[34,35].

Figure 2c shows the changes in the integrated scattering intensity around the *R-point,* induced by the resonant excitation of the highest-frequency zone-center infrared-active phonon at 17 THz. An initial increase is followed by damped oscillations about an overall long-lived reduction in the total diffuse scattering signal. These features were seen to be specific to the *R-point* in reciprocal space, as scattering at Bragg peaks and other points in the Brillouin zone only resulted in small and featureless changes. This is well exemplified by the time resolved integrated changes measured around the *M-point*, where small and slow enhancement was observed, as shown in Figure 2c.

In the following, we discuss a model for these dynamics, which inform further analysis of the data and provide a working hypothesis for light-induced ferroelectricity in STO[9]. Figure 3a shows the phonon band structure of the cubic phase of STO ($\phi = 0°$), calculated by an ab initio density functional theory (DFT) approach (see Supplementary Information for details). Amongst other phonon modes, the Brillouin zone center hosts the FE soft mode (blue), the acoustic modes related to strain (yellow), and the driven 17-THz infrared-active phonon mode (magenta). The soft phonons connected to the AFD rotation are found at the zone boundary *R-points* $q = \left(\pm\frac{1}{2} \pm \frac{1}{2} \pm \frac{1}{2}\right)$. The unstable symmetry-lowering soft modes manifest themselves as imaginary frequencies in DFT calculations, and are plotted as such. Starting from the resonant excitation of the zone-center infrared-active mode $Q_{IR}$ at frequency $\omega_{IR}$ several interaction pathways can be mapped out utilizing a density functional theory frozen-phonon approach. Three phenomena are expected.

Firstly, the driven $Q_{IR}$ mode is expected to couple nonlinearly to acoustic (strain) modes $Q_\eta$ as discussed in detail in Ref. 9. The energy of this nonlinear coupling exhibits a square-linear dependence on the driven phonon and the strain coordinate, respectively, and can be written as $V_{IR,\eta} = g_{IR,\eta} Q_{IR}^2 \cdot Q_\eta$. Also, because the pulse duration of the mid-infrared drive is short compared to the oscillation period of the strain field (determined by the ratio of the pump penetration depth and by the speed of sound), the strain wave is launched impulsively. For experimentally feasible drive electric fields, we estimate a strain wave with peak values of the order of 0.2 % [9].

Secondly, we expect a coupling of the driven mode $Q_{IR}$ to the *R-point* AFD lattice distortions $u_{AFD}$, which by symmetry has bi-quadratic character $V_{IR,AFD} = g_{IR,AFD} Q_{IR}^2 \cdot u_{AFD}^2$. This interaction potential implies a parametric excitation of pairs of phonons $u_{AFD}$ at opposite wavevectors $\pm q$ to conserve momentum. This excitation generates lattice fluctuations of the form $\langle u_{AFD}^2 \rangle$, but no net distortion $\langle u_{AFD} \rangle$ [29]. From the ab initio calculations, we find that the coupling matrix element $g_{IR,AFD}$ is negative, implying that displacements along the coordinates of $Q_{IR}$ soften the of the antiferrodistortive mode and hence enhance its fluctuations $\langle u_{AFD}^2 \rangle$, as shown in Fig. 3b.

Thirdly, we expect that the finite strain $Q_\eta$ induced at the Brillouin zone center via $V_{IR,\eta} = g_{IR,\eta} Q_{IR}^2 \cdot Q_\eta$ also couples to the *R-point* AFD distortions $u_{AFD}$. This linear-square coupling of the form $V_{\eta,AFD} = g_{\eta,AFD} Q_\eta \cdot u_{AFD}^2$ has a positive coupling coefficient $g_{\eta,AFD}$, hence hardens the antiferrodistortive mode to reduce its fluctuations $\langle u_{AFD}^2 \rangle$ (see Fig. 3c). This coupling counteracts the $V_{IR,AFD}$ interaction. Furthermore, the lifetime of the strain coupling $V_{\eta,AFD}$ to the AFD distortion is determined by the slow relaxation and propagation of the zone center acoustic phonons and is far longer lived than the $V_{IR,AFD}$ coupling, which is only significant as long as the optical phonon $Q_{IR}$ oscillates coherently.

To simulate these dynamics, we adopted the approach of Ref. 36 and calculated the time-dependent amplitude variance of the AFD distortion $\langle u_{AFD}^2[t]\rangle$ induced by the optically driven mode $Q_{IR}[t]$ and by the strain $Q_\eta[t]$ as discussed above. The coupled system of equations of motion has the following form:

$$(1) \quad \frac{\partial^2}{\partial t^2}Q_{IR}[t] + 2\gamma_{IR}\frac{\partial}{\partial t}Q_{IR}[t] + \omega_{IR}^2 Q_{IR}[t] = Z^* E[t]$$

$$(2) \quad \frac{\partial^2}{\partial t^2}Q_\eta[t] + 2\gamma_\eta \frac{\partial}{\partial t}Q_\eta[t] + \omega_\eta^2 Q_\eta[t] = g_{IR,\eta} Q_{IR}^2$$

$$(3) \quad \begin{aligned}&\frac{\partial^3}{\partial t^3}\langle u_{AFD}^2[t]\rangle + 2\gamma_{AFD}\frac{\partial^2}{\partial t^2}\langle u_{AFD}^2[t]\rangle \\ &+ 4\left(\omega_{AFD}^2 + g_{IR,AFD}\, Q_{IR}^2[t] + g_{\eta,AFD} Q_\eta[t]\right)\frac{\partial}{\partial t}\langle u_{AFD}^2[t]\rangle \\ &= -2\langle u_{AFD}^2[t]\rangle\left(g_{IR,AFD}\, \partial(Q_{IR}[t])^2/\partial t\right) + g_{\eta,AFD}\, \partial Q_\eta[t]/\partial t)\end{aligned}$$

The subscripts $IR$, $\eta$ and $AFD$ denote frequencies and lifetimes of the $Q_{IR}$ mode, the strain and the AFD distortion, respectively. $Z^*$ is the effective charge that couples the infrared-active $Q_{IR}$ mode to the external driving field, which we model as a Gaussian pulse centered at the mid-IR pump frequency: $E[t] = E_0 \sin(\omega_{IR}\, t)e^{-\frac{t^2}{2\sigma^2}}$. We determine all the coupling coefficients of these equations utilizing a first principles approach based on DFT and adjust a few of them to best match the experimental results (see Supplementary Information for details).

In Fig. 4 we compare the simulated variance $\langle u_{AFD}^2[t]\rangle$ to the experimentally determined time-dependent *R-point* scattering intensities, measured for different excitation fluences. From the simulation, we can isolate the dynamics arising from the different interactions. First, the $V_{IR,AFD}$ coupling alone (red curve) results in a picosecond-lived oscillation of $\langle u_{AFD}^2[t]\rangle$ at twice the AFD soft mode frequency as the result of mode squeezing of this mode in combination with the short rise time of the $Q_{IR}$ mode oscillations. The negative sign of the coefficient $g_{IR,AFD}$ leads to a softening of the potential of the AFD distortion, resulting in the initial increase of the variance $\langle u_{AFD}^2[t]\rangle$. Next, we simulate the fluctuations of the *R-point* AFD distortion when coupled only to the optically induced strain via $V_{\eta,AFD}$ (orange curve). In this case, a slow monotonic decrease of $\langle u_{AFD}^2[t]\rangle$ is observed because the induced strain squeezes the AFD soft mode with a rise time that is too slow to launch oscillations in its amplitude variance. Note that the slow enhancement of the scattering intensity at the *M-point,* shown in Fig. 2c, is of the same origin but with a coupling coefficient of opposite sign (see Supplementary Information for details).

Taken together, the time-dependent amplitude of the AFD lattice fluctuations $\langle u_{AFD}^2[t]\rangle$ is driven by a short-lived phonon-phonon interaction and a longer-lasting strain-phonon interaction. Importantly, the instantaneous shape of the AFD potential also determines the oscillation frequency of the variance $\langle u_{AFD}^2[t]\rangle$. Consequently, a larger strain induced by a higher amplitude of the driven $Q_{IR}$ phonon results in higher-frequency $\langle u_{AFD}^2[t]\rangle$ oscillations. This expectation is experimentally confirmed in the excitation fluence measurements shown in Fig. 4c,e and reproduced in the corresponding simulations in Figs. 4d,f.

Having established the antiferrodistortive dynamics driven in the cubic STO phase, we now discuss their implications on the light-induced ferroelectricity found in Ref. 9. Our study shows that driving the $Q_{IR}$ phonon mode produces short-lived oscillations and initial enhancement of the *R-point* AFD fluctuations $\langle u_{AFD}^2[t]\rangle$, but at longer times it creates a state in which these fluctuations are suppressed. Figure 5a shows the corresponding modifications in the energy potential of the AFD distortion. The red areas indicate the spread of the fluctuations $\langle u_{AFD}^2[t]\rangle$, which act dynamically also on the FE distortions in the cubic ground state. This is illustrated in Figure 5b that is a zoom into Fig. 1c around $\phi = 0$. At negative time delay, fluctuations of the AFD distortions $\langle u_{AFD}^2[t]\rangle$ cover a certain width in rotation angle $\phi$. These fluctuations are expected to suppress the ferroelectric state, because AFD rotations away from $\phi = 0$ reduce the depth of the potential energy along the coordinate of the ferroelectric distortion. Then, excitation of the infrared-active phonon and its bi-quadratic coupling to the *R-point* AFD soft mode enhances the fluctuations $\langle u_{AFD}^2[t] \rangle$ to suppress the ferroelectric state even more. However, the onset of the strain distortion at long times after the excitation sizably reduces the AFD fluctuations, so that the phase space, occupied in the FE energy surface, becomes even smaller than at equilibrium. In this situation the condensation of the FE state becomes more likely, and may explain the growth of the FE state. Note that in the experiments reported in Ref. 9, a light-induced ferroelectric phase was observed up to room temperature, that is both below and above the equilibrium transition into the tetragonal phase. The discussion above only applies to the high-temperature regime (T > 110 K), in which paraelectric SrTiO$_3$ is cubic. For lower temperatures, we expect the same elements discussed above to still be valid, although additional effects could contribute to amplify the photo-induced state.

In summary, we have used time resolved diffuse x-ray scattering and ab-initio DFT simulations to clarify the physics of photo-induced ferroelectricity in SrTiO$_3$. We show results that go beyond the measurements of the average position of the atoms in the unit cell captured by Bragg diffraction. We also show how large fourth-order lattice interactions affect the functional response of materials. We expect that in the magnetically ordered fluoride perovskite KMnF$_3$, which has many common features with the STO, modifications of octahedral rotations at the (½ ½ ½) *R-point* may significantly affect the exchange interactions between *d*-electrons of the Mn$^{2+}$ cations whose spins antiferromagnetically order at the same wave vector[37]. Control of high-order phonon interactions is in fact a frontier in the use of nonlinear phononics to manipulate the functional properties of solids[8].

**Acknowledgements:**

The research leading to these results received funding from the Deutsche Forschungsgemeinschaft (German Research Foundation) via the excellence cluster "CUI: Advanced Imaging of Matter" (EXC 2056, project ID 390715994). G. O., V. K., G dlP., Q. L N. and M.T. were supported by the U.S. Department of Energy, Office of Science, Office of Basic Energy Sciences through the Division of

**Figures:**

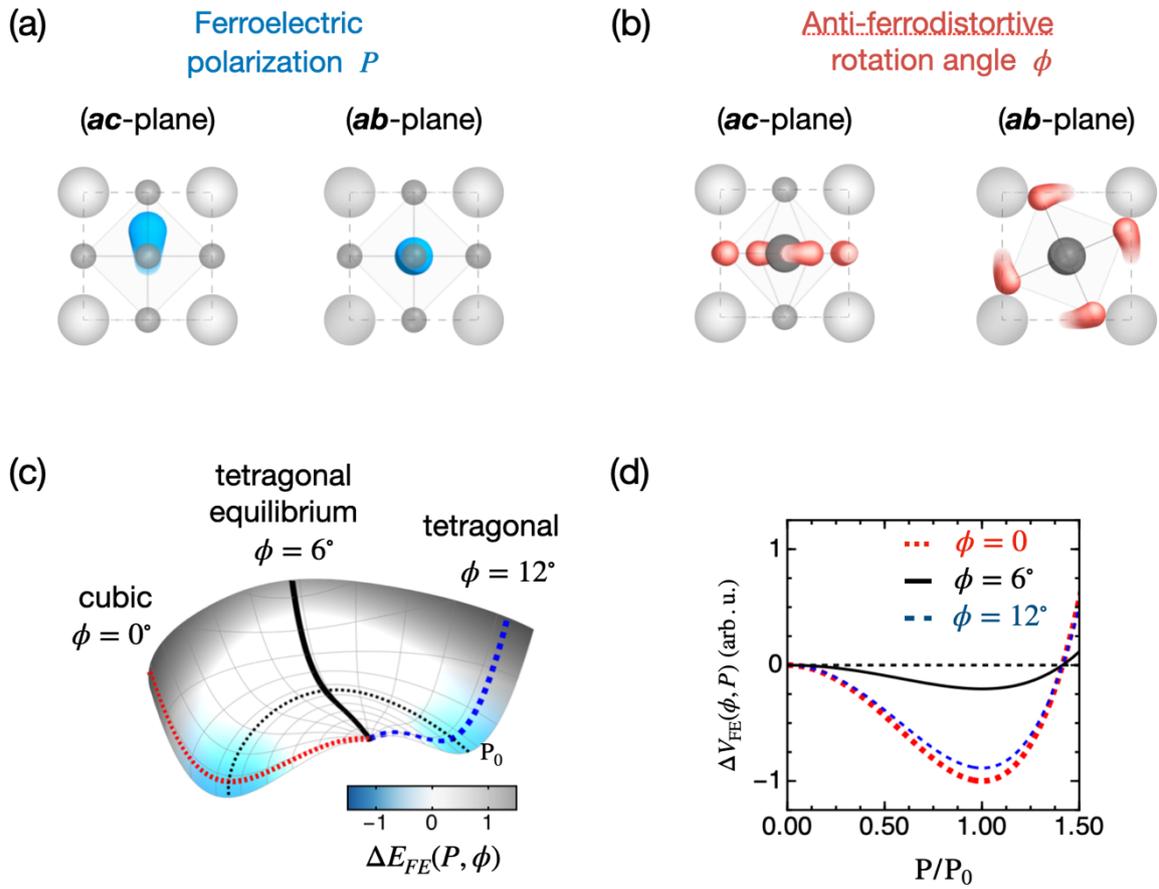

*Figure 1. | **The fundamental distortions of SrTiO₃.** (a) The polar distortion, creating the symmetry-broken ferroelectric state with polarization P, involves the displacement of the center Ti atom along the c-axis. (b) The antiferrodistortive distortion involves the rotation of the oxygen octahedron around the c axis by an angle ϕ. (c) Illustration of the energy landscape of the ferroelectric and the antiferrodistortive distortions. The minimum in ferroelectric energy $\Delta E_{FE}$ of the cubic phase ($\phi = 0$) reduces when the antiferrodistortive rotation sets in. At the rotation angle of the tetragonal equilibrium state ($\phi = 6°$) the energy surface exhibits a saddle point where $\Delta E_{FE}$ is minimized. (c) Selected cuts through this energy surface.*

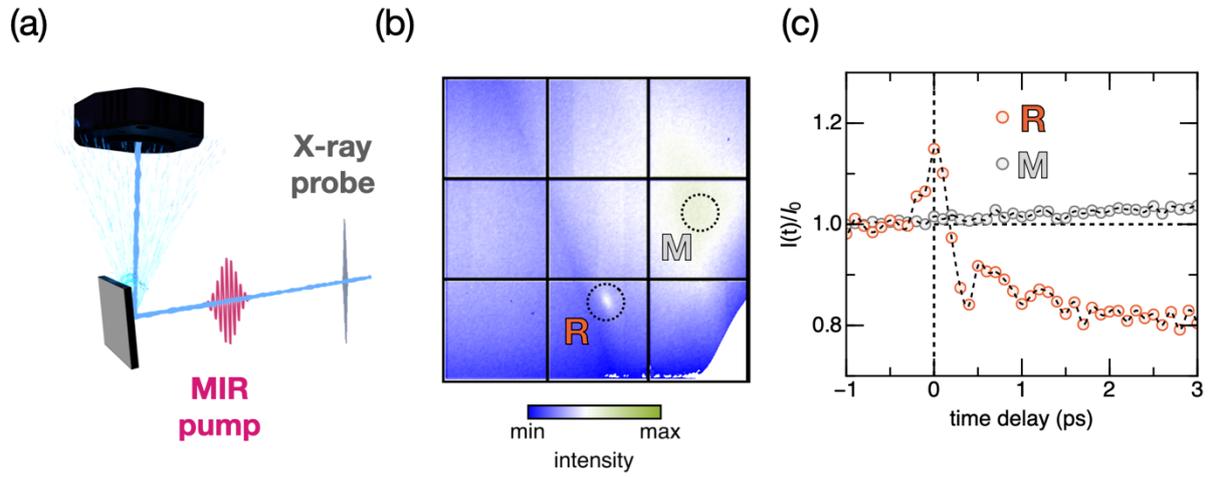

***Fig. 2 | Time-resolved x-ray diffuse scattering.*** *(a) An intense mid-infrared pulse resonantly excites the highest-frequency SrTiO$_3$ IR-active phonon mode. Diffraction of a time-delayed femtosecond x-ray pulse probes the resulting lattice dynamics in reciprocal space. (b) Two-dimensional x-ray detector image with selected high-symmetry points **R**- (1/2,1/2,1/2) and **M** (1/2,1/2,0.0) at equilibrium. The R-point hosts the antiferrodistortive fluctuations of the cubic-to-tetragonal phase transition. (c) Measured changes in x-ray scattering intensity at the R- and M-point induced by the nonlinear excitation of the crystal lattice.*

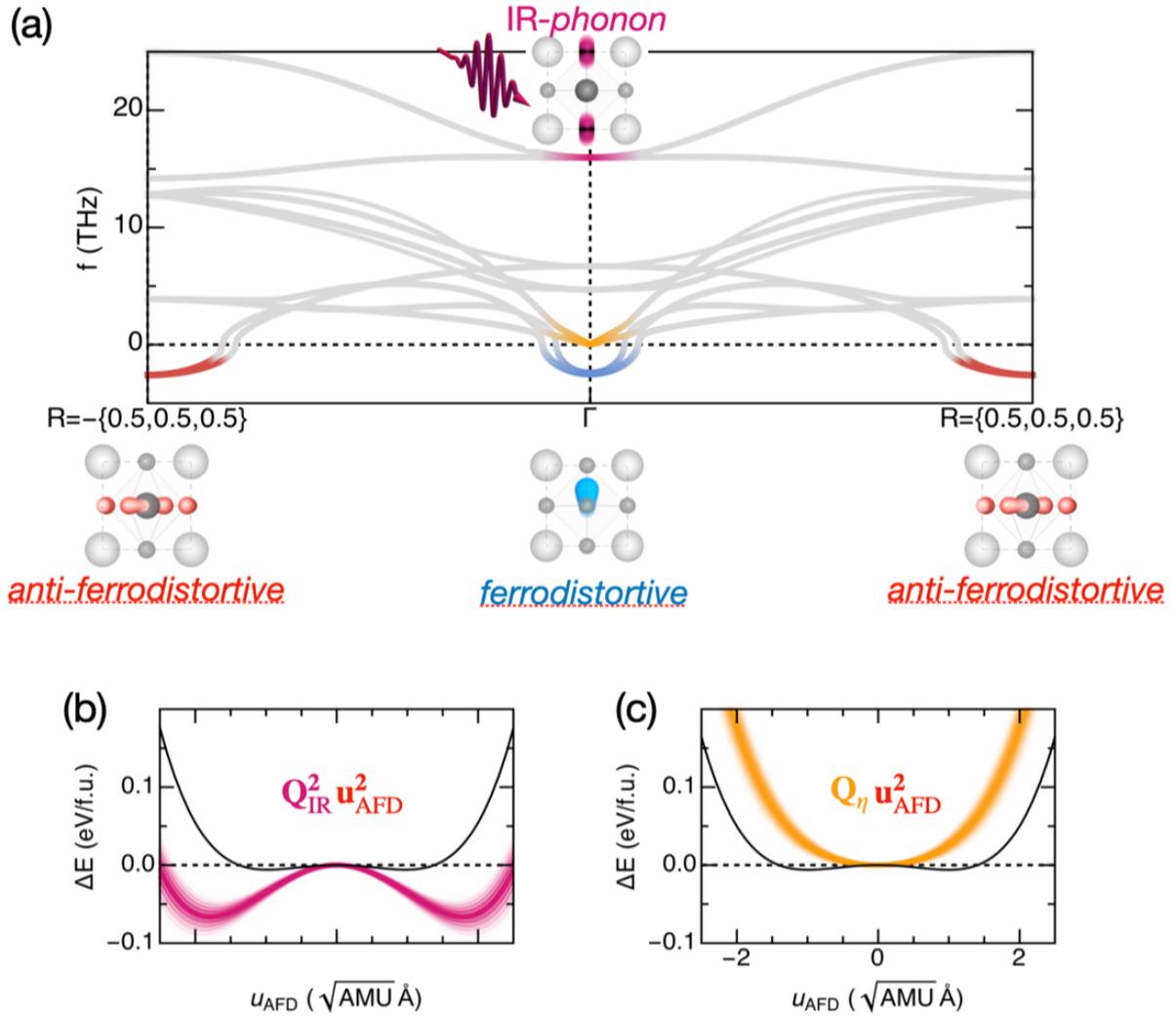

**Fig. 3 | Anharmonic phonon-phonon coupling across the Brillouin zone.** (a) Phonon dispersion of cubic ($\phi = 0$) SrTiO$_3$ along the $R - \Gamma - R$ direction calculated from an ab-initio approach. Colored lines highlight the positions of the driven infrared-active phonon (magenta), strain waves (orange), the ferroelectric soft mode (blue), all at the Brillouin zone center, and of the antiferrodistortive mode (red) at the zone edge. Negative frequencies represent unstable (soft) phonon modes. (b) and (c) Energy potentials of the R-point antiferrodistortive distortion as a function of amplitude $u_{AFD}$. Black lines are the equilibrium potential. Colored lines show modifications of this potential due to nonlinear coupling to the zone-center infrared-active mode $Q_{IR}$ and strain $Q_\eta$, as discussed in the text.

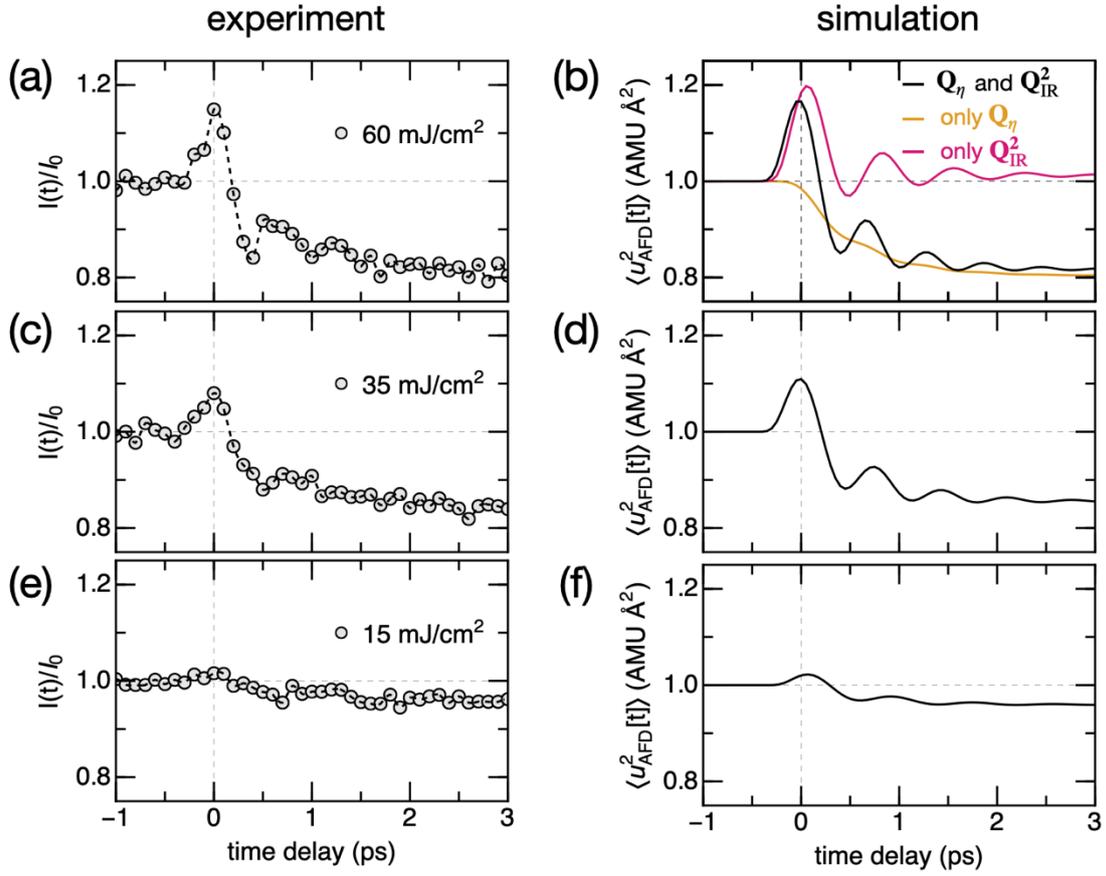

**Fig. 4 | Fluence dependent measurements and simulations.** (a,c,e) Time-resolved changes of the R-point x-ray scattering intensity for different mid-infrared excitation fluences. The initial rise of the scattering intensity is followed by oscillations with a fluence dependent frequency and a slow decrease. (b,d,f), Simulations of the variance of the antiferrodistortive R-point amplitude $\langle u^2_{AFD}[t]\rangle$ utilizing the model presented in Eqs. (1-3). The contributions of its coupling to the infrared-active phonon $Q_{IR}$ (magenta) and strain $Q_\eta$ (orange) are individually shown in panel b (see the corresponding non-equilibrium potentials in Fig. 3b,c).

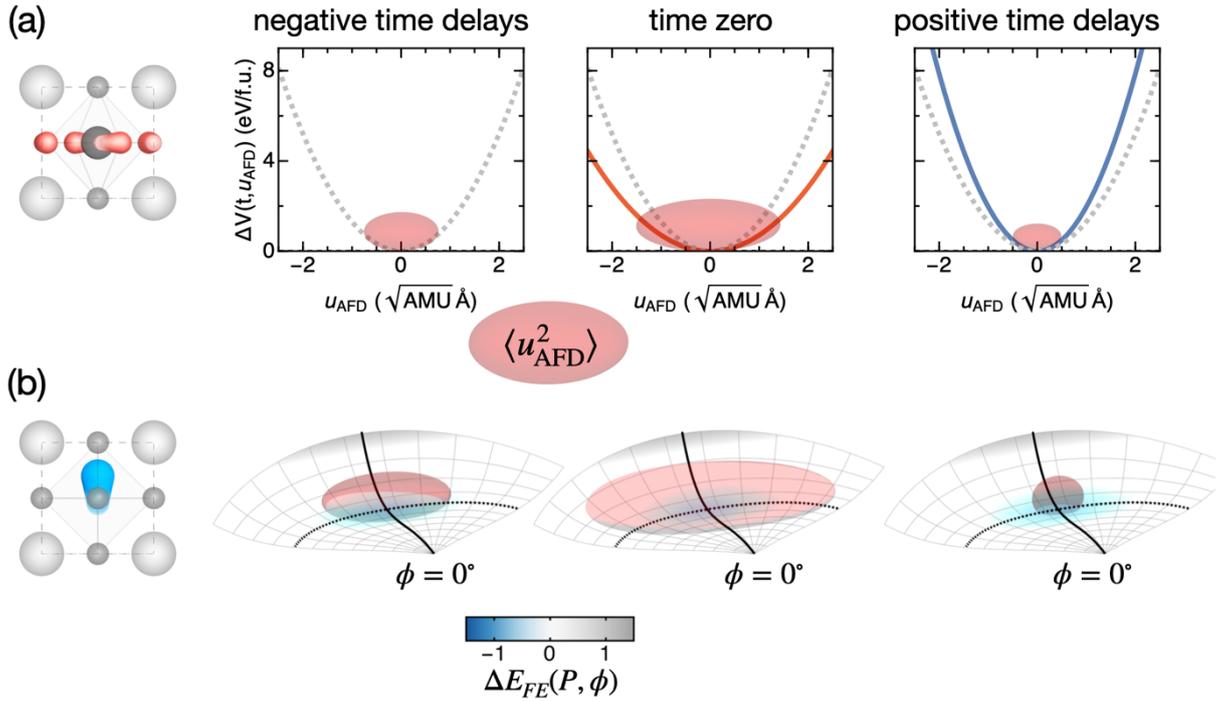

**Fig 5. | Impact** *of the antiferrodistortive lattice fluctuations* **on the ferroelectric energy gain.** *(a) Potential of the antiferrodistortive mode at equilibrium (grey, negative time delay) the corresponding amplitude variance indicated as red shaded area. At zero time delay, the potential softens due to coupling to the resonantly driven infrared-active phonon mod, hence the phase space of the fluctuations $\langle u_{AFD}^2[t] \rangle$ is enhanced. At longer times, the onset of strain hardens the potential, thereby reducing $\langle u_{AFD}^2 \rangle$ fluctuations to below the equilibrium value. (b), Impact of these dynamics on the ferroelectric energy landscape. At negative time delays, the thermal fluctuations $\langle u_{AFD}^2 \rangle$ cover the entire region in which the ferroelectric state gains energy, hence prohibiting a condensation of this mode. This behavior becomes pronounced at zero time delay, where the fluctuations are enhanced. At longer times, the reduced $\langle u_{AFD}^2 \rangle$ fluctuations cover only the small area where the ferroelectric state gains energy, enabling a condensation of the ferroelectric soft mode.*

**References:**

1.  Zhang, J. & Averitt, R. D. Dynamics and Control in Complex Transition Metal Oxides. *Annu Rev Mater Res* **44**, 19–43 (2014).

2.  Basov, D. N., Averitt, R. D. & Hsieh, D. Towards properties on demand in quantum materials. *Nat Mater* **16**, 1077–1088 (2017).

3.  Mankowsky, R., von Hoegen, A., Först, M. & Cavalleri, A. Ultrafast Reversal of the Ferroelectric Polarization. *Phys Rev Lett* **118**, 197601 (2017).

4.  Juraschek, D. M., Fechner, M., Balatsky, A. v. & Spaldin, N. A. Dynamical multiferroicity. *Phys Rev Mater* **1**, 014401 (2017).

5.  Radaelli, P. G. Breaking symmetry with light: Ultrafast ferroelectricity and magnetism from three-phonon coupling. *Phys Rev B* **97**, 085145 (2018).

6.  Disa, A. S. *et al.* Polarizing an antiferromagnet by optical engineering of the crystal field. *Nat Phys* **16**, 937–941 (2020).

7.  Stupakiewicz, A. *et al.* Ultrafast phononic switching of magnetization. *Nat Phys* **17**, 489–492 (2021).

8.  Disa, A. S., Nova, T. F. & Cavalleri, A. Engineering crystal structures with light. *Nat Phys* **17**, 1087–1092 (2021).

9.  Nova, T. F., Disa, A. S., Fechner, M. & Cavalleri, A. Metastable ferroelectricity in optically strained SrTiO3. *Science* **364**, 1075–1079 (2019).

10. Li, X. *et al.* Terahertz field–induced ferroelectricity in quantum paraelectric SrTiO3. *Science* **364**, 1079–1082 (2019).

11. Henstridge, M., Först, M., Rowe, E., Fechner, M. & Cavalleri, A. Nonlocal nonlinear phononics. *Nat Phys* **18**, 457–461 (2022).

12. Kuzmanovski, D., Aeppli, G., Rønnow, H. M. & Balatsky, A. v. Kapitza stabilization of a quantum critical order. *arXiv:2208.09491* (2022).

13. Zhuang, Z., Chakraborty, A., Chandra, P., Coleman, P. & Volkov, P. A. Light-Driven Transitions in Quantum Paraelectrics. *arXiv:2301.06161* (2023).

14. Sokolowski-Tinten, K. *et al.* Femtosecond X-ray measurement of coherent lattice vibrations near the Lindemann stability limit. *Nature* **422**, 287–289 (2003).

15. Mankowsky, R. *et al.* Nonlinear lattice dynamics as a basis for enhanced superconductivity in YBa2Cu3O6.5. *Nature* **516**, 71–73 (2014).

16. Huber, T. *et al.* Coherent structural dynamics of a prototypical charge-density-wave-to-metal transition. *Phys Rev Lett* **113**, 026401 (2014).

17. Gerber, S. *et al.* Femtosecond electron-phonon lock-in by photoemission and x-ray free-electron laser. *Science* **357**, 71–75 (2017).

18. Buzzi, M., Först, M., Mankowsky, R. & Cavalleri, A. Probing dynamics in quantum materials with femtosecond X-rays. *Nat Rev Mater* **3**, 299–311 (2018).

19. Kozina, M. *et al.* Terahertz-driven phonon upconversion in SrTiO3. *Nat Phys* **15**, 387–392 (2019).

20. Porer, M. *et al.* Ultrafast transient increase of oxygen octahedral rotations in a perovskite. *Phys Rev Res* **1**, 012005 (2019).

21. Casals, B. *et al.* Low-Temperature Dielectric Anisotropy Driven by an Antiferroelectric Mode in SrTiO3. *Phys Rev Lett* **120**, 217601 (2018).

22. Wall, S. *et al.* Ultrafast disordering of vanadium dimers in photoexcited VO2. *Science* **362**, 572–576 (2018).

23. Müller, K. A. & Burkard, H. SrTiO3: An intrinsic quantum paraelectric below 4 K. *Phys Rev B* **19**, 3593–3602 (1979).

24. Rytz, D., Höchli, U. T. & Bilz, H. Dielectric susceptibility in quantum ferroelectrics. *Phys Rev B* **22**, 359–364 (1980).

25. Itoh, M. *et al.* Ferroelectricity Induced by Oxygen Isotope Exchange in Strontium Titanate Perovskite. *Phys Rev Lett* **82**, 3540–3543 (1999).

26. Mitsui T. & Westphal W. B. Dielectric and X-Ray Studies of CaxBa1–xTiO3 and CaxSr1–xTiO3. *Physical Review* **124**, 1354–1359 (1961).

27. Haeni, J. H. *et al.* Room-temperature ferroelectricity in strained SrTiO3. *Nature* **430**, 758–761 (2004).

28. Aschauer, U. & Spaldin, N. A. Competition and cooperation between antiferrodistortive and ferroelectric instabilities in the model perovskite SrTiO3. *Journal of Physics: Condensed Matter* **26**, 122203 (2014).

29. Trigo, M. *et al.* Fourier-transform inelastic X-ray scattering from time- and momentum-dependent phonon–phonon correlations. *Nat Phys* **9**, 790–794 (2013).

30. Ingold, G. *et al.* Experimental station bernina at swissfel: Condensed matter physics on femtosecond time scales investigated by x-ray diffraction and spectroscopic methods. *J Synchrotron Radiat* **26**, 874–886 (2019).

31. B.E. Warren. *X-Ray Diffraction (Dover Books on Physics)*. (1990).

32. Yamada, Y. X-Ray critical diffuse scattering at a structural phase transition. *Ferroelectrics* **7**, 37–43 (1974).

33. Shirane, G. & Yamada, Y. Lattice-Dynamical Study of the 110°K Phase Transition in SrTiO3. *Physical Review* **177**, 858–863 (1969).

34. Andrews, S. R. X-ray scattering study of the R-point instability in SrTiO3. *Journal of Physics C: Solid State Physics* **19**, 3721–3743 (1986).

35. Darlington, C. N. W. & O'Connor, D. A. The central mode in the critical scattering of X-rays by SrTiO3. *Journal of Physics C: Solid State Physics* **9**, 3561–3571 (1976).

36. Garrett, G. A., Whitaker, J. F., Sood, A. K. & Merlin, R. Ultrafast optical excitation of a combined coherent-squeezed phonon field in SrTiO3. *Opt Express* **1**, 385 (1997).

37. Heeger, A. J., Beckman, O. & Portis, A. M. Magnetic Properties of KMnF3. II . Weak Ferromagnetism. *Physical Review* **123**, 1652–1660 (1961).